# An Electrochemical Potentiostat Interface for Mobile Devices: Enabling Remote Medical Diagnostics

Henry Fu, Henry Chow, Michael Lew, Shruti Menon, Craig Scratchley, M. Ash Parameswaran

## Abstract

An electrochemical potentiostat interface for mobile devices has been designed and implemented. The interface consists of a potentiostat module, a microcontroller module, and a Bluetooth module. The potentiostat module performs electrochemical measurements and detects the responses from the samples. The microcontroller module controls the test and communication processes. The Bluetooth module links the system to a mobile device, where the mobile device acts as a control-console, data storage system, communication unit, and graphical plotter for the overall diagnostic processes. This interface is suitable for point-of-care and remote diagnostics, enhancing the capabilities of mobile devices in telemedicine.

## Keywords

Mobile device; Remote medical diagnostics; Potentiostat, Bluetooth; Cyclic voltammetry; Electrochemical detection

## 1. Introduction

Intelligent, portable electronic systems can assist healthcare professionals to efficiently perform medical diagnostics [1-4]. This is particularly important in remote areas and developing countries, where the lack of local medical facilities and personnel could delay diagnosis and treatment. Timely diagnosis is critical especially when dealing with newborn infants, whereas a delay may lead to an increase in the severity of condition or even cause death. To address this issue, there is a growing interest to develop remote and onsite diagnostic technologies and instruments. Ideally, operation of the remote diagnostic equipment should require minimal or no training. An intuitive user interface should allow the user to perform diagnostics easily using the mobile device. This approach opens up the possibility to incorporate intelligent algorithms in the mobile device for an automated diagnostic advisory as well as the ability to upload the test-data, through a telecommunication link, to a central location where a professional can examine the results and offer diagnostic consultancy and treatment plan.

High-resolution displays, touch-screen input, Bluetooth and WiFi connectivity, USB ports, GPS, and many other sensors are becoming ubiquitous even in simple mobile devices. Integration of a diagnostic system using these mobile devices generates new implementation possibilities revolutionizing the power of telemedicine.

## 2. Potentiostat Interface

Electrochemistry is one of the powerful emerging analytical tools in medical diagnostics [5-7]. Literature shows electrochemical tests for diagnosing cancer [5], cardiovascular diseases [6] and antibiotic susceptibility [7]. Typically, an electrochemical test requires an electronic-circuit called a potentiostat. A potentiostat is an electronic device that is connected to a 3-electrode-unit (sensor) which is immersed in a biochemical (body-fluid) sample. The potentiostat applies a voltage and records the current response that is characteristic of the sample being tested. This response is used in diagnostics. Integration of the potentiostat to a mobile device makes the diagnostic unit portable. To interface the potentiostat to a mobile device, a communication link is required. Bluetooth was chosen for this purpose as it features both low power and wireless communication. The latter allows the mobile platform to be isolated from the test sample. Additionally, Bluetooth modules are small in size and are ideally suited for portable systems. The potentiostat interface includes a microcontroller for managing the functionality. To ensure portability the potentiostat interface should be compact and consume minimal electrical power. The WiFi or cellular module of mobile devices can be used to transfer the test-results to medical professionals for analysis.

## 3. Implementation

### 3.1. System Overview

A photograph of the developed prototype Mobile Device Interface (MDI) is shown in Figure 1. The block diagram of the MDI is shown in Figure 2. The MDI consists of the following units: potentiostat module; Bluetooth module (Panasonic CC2560-PAN1325); microcontroller (TI MSP430BT5190); power (battery); and test electrode connectors. The microcontroller operates the potentiostat to perform the electrochemical measurements and acquires the data. This data is then transmitted to the mobile device via Bluetooth.

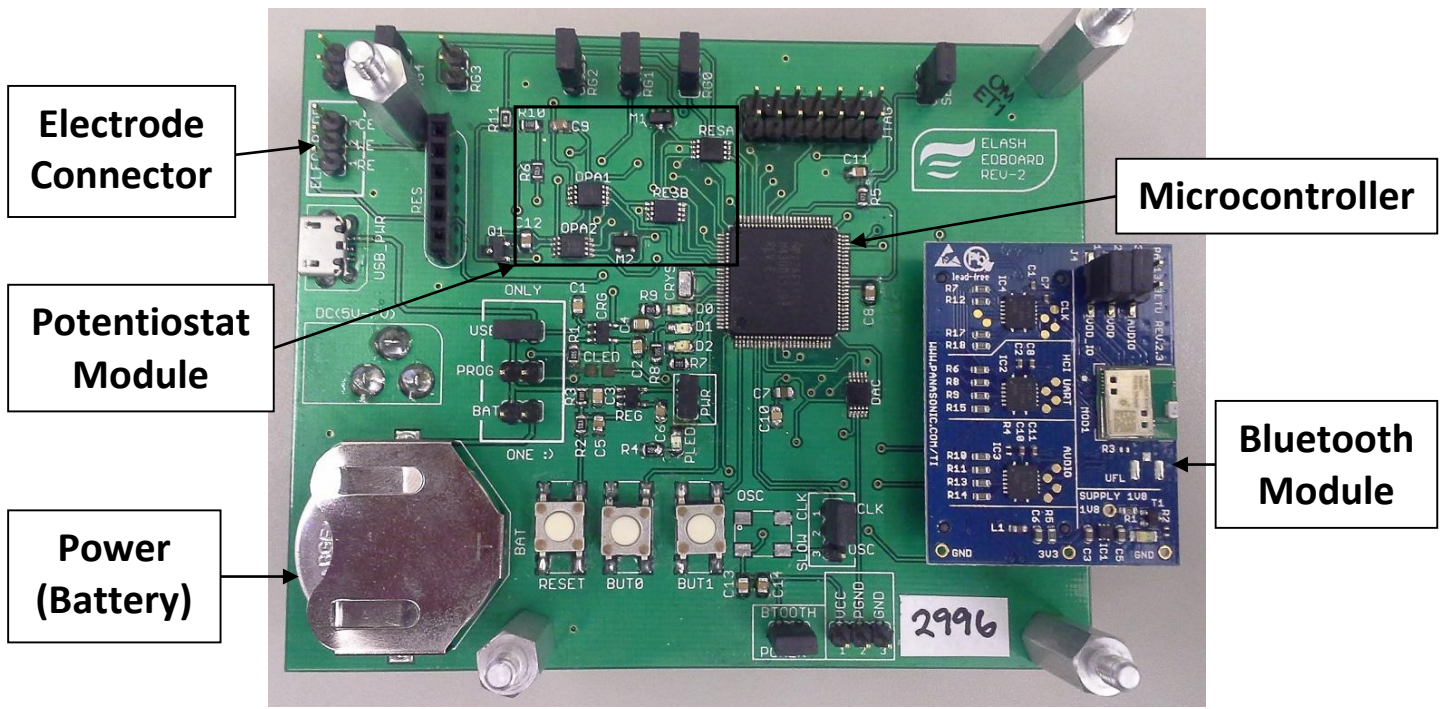

Figure 1. The Mobile Device Interface prototype. Total dimensions 80mm x 110mm.

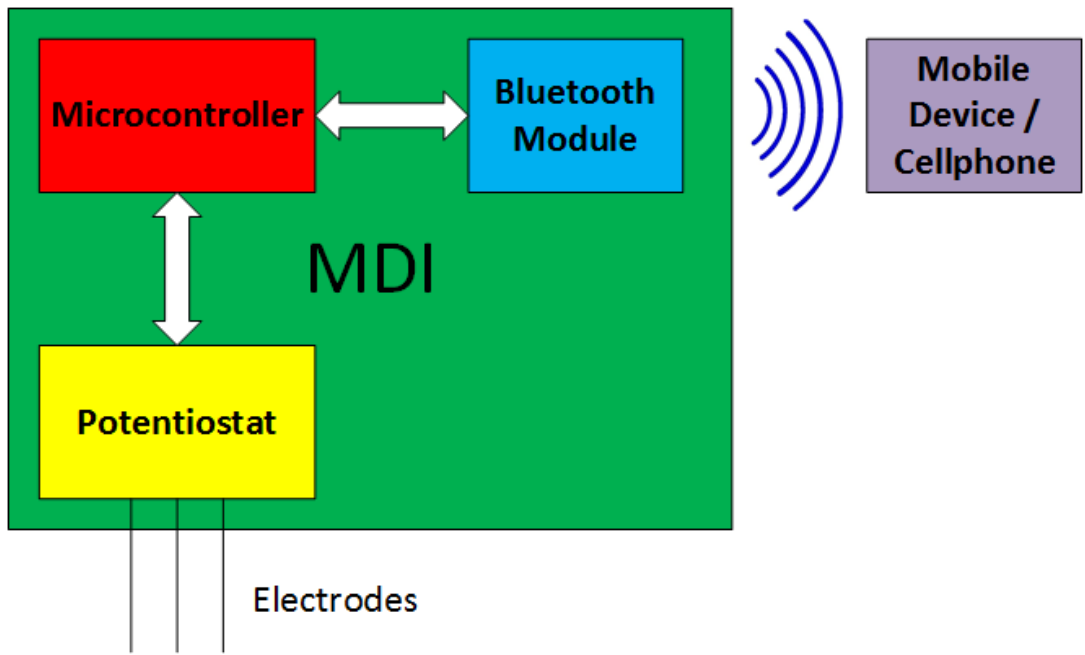

Figure 2. High Level Block Diagram of MDI.

In typical operation, the test parameters determined by the user are sent to the MDI through the mobile device. The MDI performs the tests and returns the results to the mobile device. The transaction between the MDI and the mobile device is through Bluetooth. In the MDI, the communication between the microcontroller and Bluetooth is through an asynchronous link. A progress bar is displayed on the mobile device when the test is in progress. After the test is completed, the electrochemical response is plotted.

## 3.2. Potentiostat

A potentiostat is an electronic circuit that performs electrochemical tests using electrodes on a biochemical sample [8]. A standard electrochemical test uses three electrodes: Working Electrode (WE), Reference Electrode (RE) and a Counter Electrode (CE), see Figure 3. The WE must be good electronic conductor and electrochemically inert as it exchanges electrons at the electrode-solution interface. Common WEs are platinum, gold, mercury and the glassy carbon electrode (GCE). The reaction at the WE occurs with respect to the RE. Thus a good RE should maintain its potential for small currents. Potential values are typically reported with respect to the RE. Common REs are the saturated calomel electrode and the silver/silver chloride (Ag/AgCl) electrode. The CE can be any electrode that does not produce reaction products that would interfere with the reaction. Its purpose is to balance out the redox reaction occurring at the WE. Common CEs are made from electrochemically inert materials such as platinum, gold and carbon [9].

There are a number of scanning techniques that are used in electrochemical tests such as, Cyclic Voltammetry (CV), Linear Sweep Voltammetry (LSV) and Square Wave Voltammetry (SWV) [10]. CV was used for the tests presented in this paper as it is the most popular scanning technique in electrochemistry. In CV, the input signal is a triangular waveform as illustrated in Figure 4a. The input triangular waveform signal probes the electrochemical signature of the sample. If the sample contains electrochemically active molecules, a redox reaction is initiated which returns a characteristic signature in the output as illustrated in Figure 4b. The redox reaction occurs at specific voltages, and this is what is used for electrochemical analysis and eventual diagnostics.

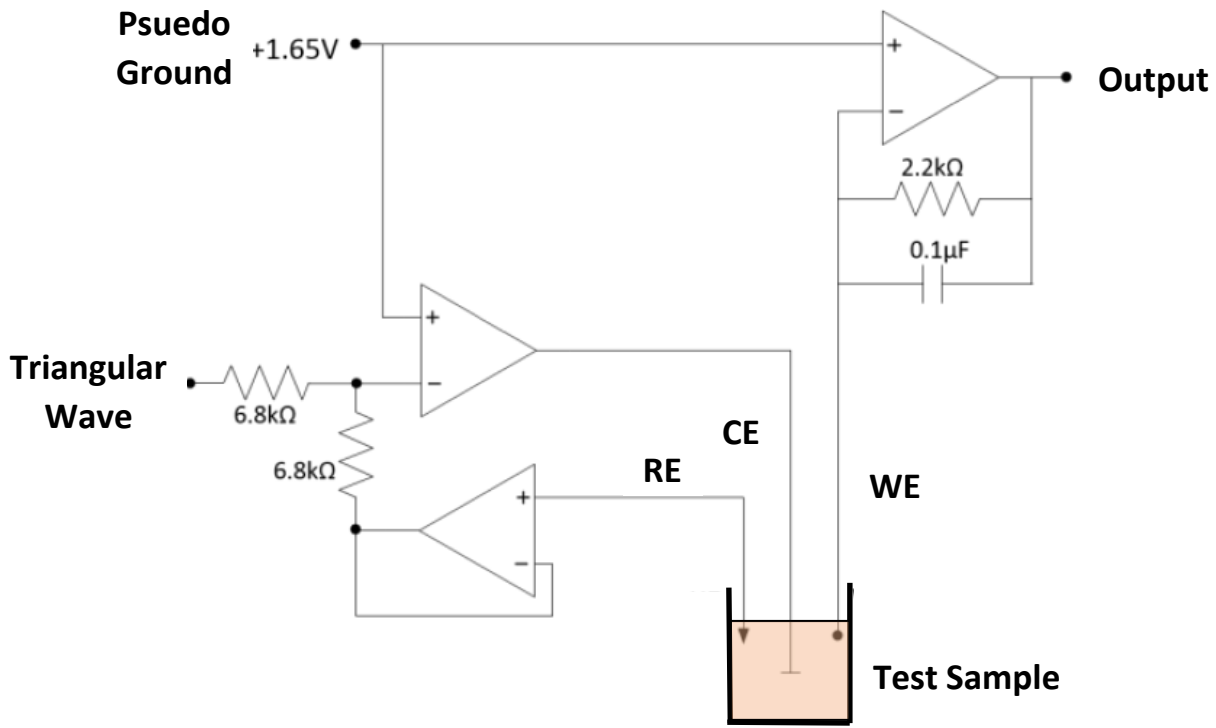
Figure 3. Potentiostat Schematic.

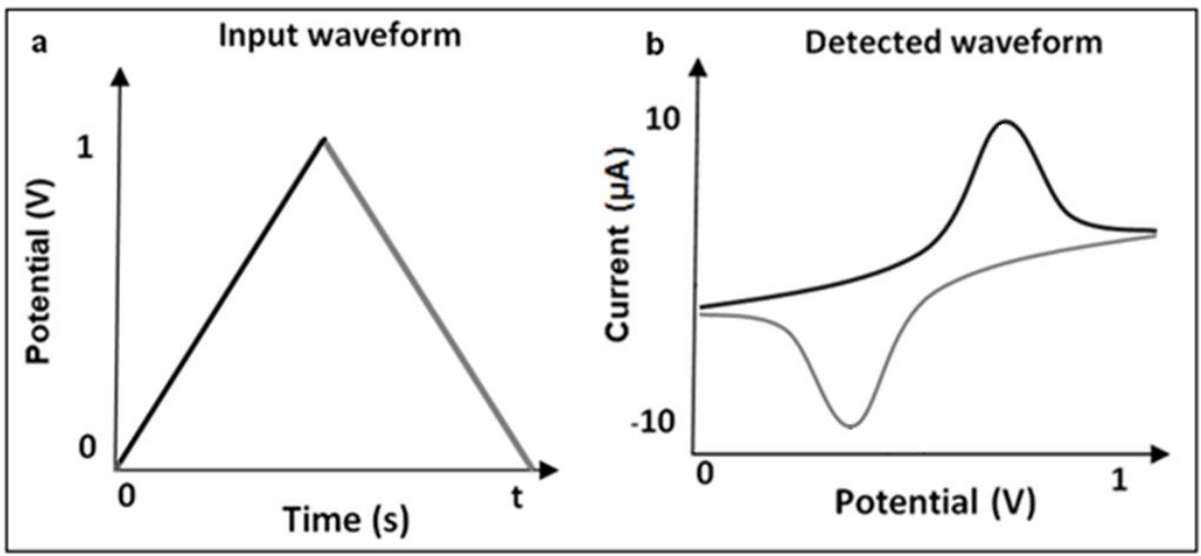
Figure 4. [a] Cyclic voltammetry input waveform [b] and an example of a resulting output plot.

## 4. MDI Software

The following subsections describe the software development for the MDI. The potentiostat is an analog circuit that generates the input waveform used to test the sample. The microcontroller samples the input and output waveforms and uses the Bluetooth module, connected via UART, to send the data to the smartphone.

### 4.1. Bluetopia®

Bluetopia® is a Bluetooth protocol stack developed by Stonestree One, LLC that supports the MDI hardware platform. This stack is the hardware abstraction layer for the Bluetooth hardware. In this implementation, the Serial Port Profile (SPP) is utilized to emulate an RS-232 serial cable connection between two peer devices using Radio Frequency Communications (RFCOMM) as the underlying protocol [11].

### 4.2. Software Architecture

The kernel supports non-preemptive multithreading using a round robin scheduling algorithm. The first thread is dedicated to controlling the Bluetooth module using the Bluetooth stack to periodically check for incoming commands from the smartphone and to send data back while the test is in progress. The second thread controls the potentiostat circuit by toggling GPIO pins and measuring the voltage using the on-chip analog to digital converter. Data is passed between the two threads using simple queues. Since the microcontroller has only one core and the kernel is non-preemptive, there is no need to guard against potential synchronization problems when using these queues.

### 4.3. Handling Commands from the Mobile Device

There are only two valid commands the device will accept from the mobile device: "start scan" and "abort". The start command contains the user defined scan parameters and the microcontroller will initiate the scan and begin sending data back to the mobile device. The abort command cancels any ongoing scan and resets the potentiostat circuit back to its idle state.

### 5. Android Application

To implement a low-cost portable diagnostic system using the developed MDI requires a Bluetooth-capable mobile platform, as mentioned previously, and programmability. Android satisfies these requirements, it is the most popular platform and is easy to work with. As literature reports, Android is currently the most popular smartphone operating system with 76.6% market share in the fourth-quarter of 2014 [12].

The Android application consists of two primary modules:

- Graphical User Interface (GUI)
- Bluetooth Communication Management

### 5.1. Graphical User Interface

The GUI includes features for setup for the test scan, save of scan results, load of previous scan results, and graphical display of test results. The GUI contains three major sections: *main* menu (with settings), *load test* menu, and *graph results*.

### 5.1.1. Main Menu

Figure 5 shows the screenshot of the *main menu*. At the beginning of a test, the user is prompted to input a Test Name. The Scan Settings input fields will allow configuration of data acquisition dependencies of the MDI. The input fields include parameters such as: slope, starting voltage, minimum voltage, maximum voltage, number of cycles, and inversion. All Scan Setting values can be modified on the mobile device. Only numerical values are allowed for input, except for the inversion option, which is a toggle.

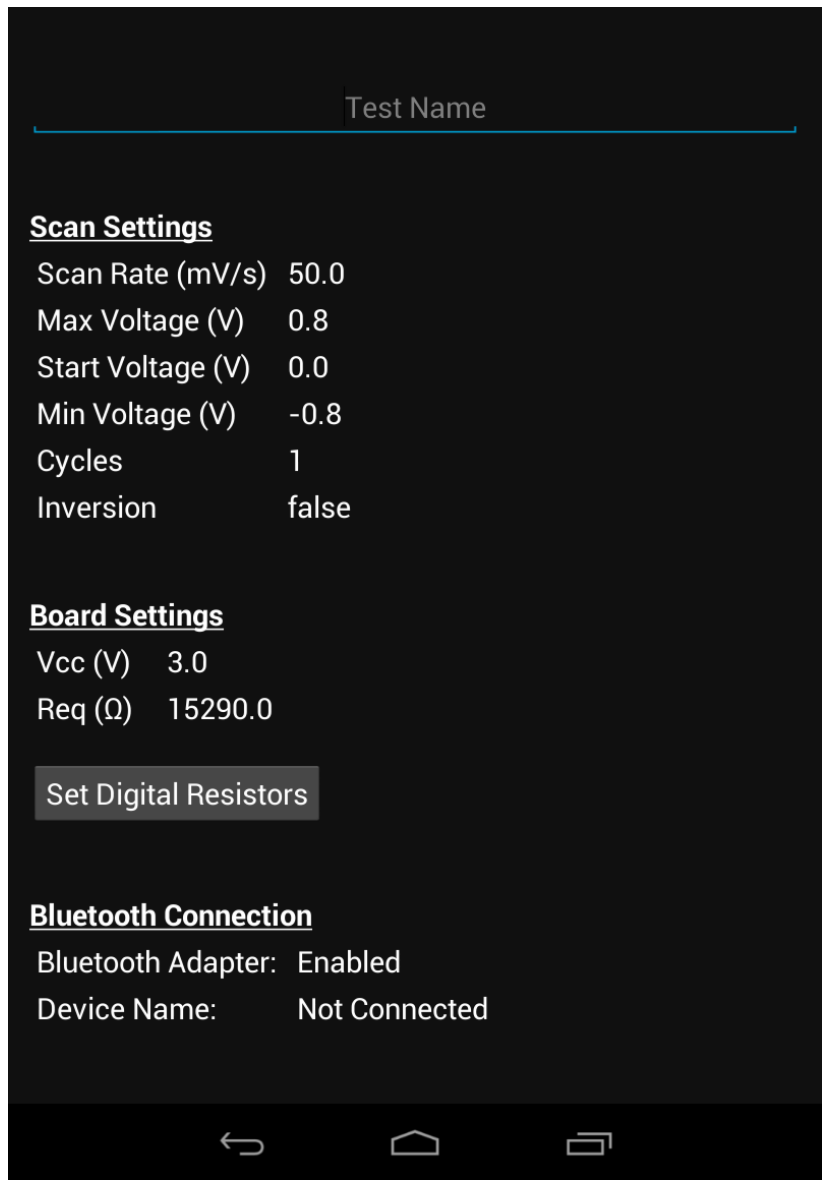

Figure 5. Screenshot of Main Menu.

The numerical input settings are programmed to have a limit so that the user cannot input out-of-range values.

Table 1 lists the permissible Scan Setting ranges.

Table 1. Allowable Values for Test Parameters.

| Scan Settings | Symbol | Unit | Minimum Value | Maximum Value | Resolution | Default Value |
|---|---|---|---|---|---|---|
| Scan Rate | - | mV/s | 1.0 | 100.0 | 1.0 | 50.0 |
| Min Voltage | $V_{min}$ | V | 0.30 | $V_{max} - 0.10$ | 0.50 | 0.30 |
| Max Voltage | $V_{max}$ | V | $V_{min} - 0.10$ | 3.0 | 0.05 | 3.00 |
| Start Voltage | - | V | $V_{min} + 0.10$ | $V_{max} - 0.10$ | 0.05 | 1.65 |
| Number of Cycles | - | - | 1 | 9999 | 1 | 1 |

### 5.1.2. Load Test Menu

The *load test* menu will allow the user to access previous test results. It will display a list of the file names of the previous tests. The test data is stored in Comma Separated Values (CSV) format. The file names contain a time stamp as well as a Test Name. The files are sorted according to the time stamp, with the newest file on the top. The user can plot the test result data by tapping on the file name. The user can also delete a specific file by long press and confirmed when prompted. A notification will appear showing the name of the deleted file.

### 5.1.3. Graph Results

This option will process the collected test data and plot it on the screen. The user can zoom or pan the display. Figure 6 shows a screenshot of a test data plot. This plotting is implemented based on AChartEngine [13], which is an open source software library for data plotting.

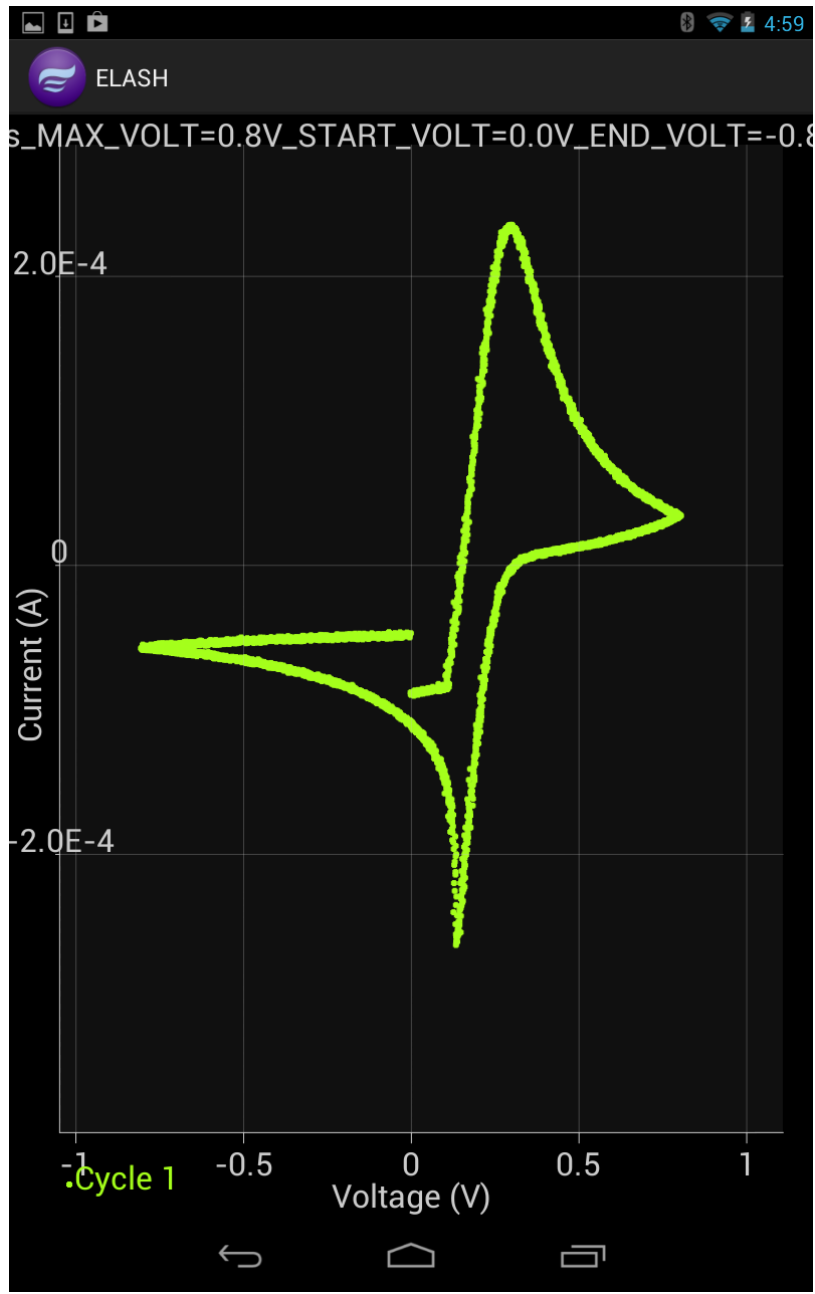

**Figure 6. Screenshot of CV Plot.**

### 5.2. Bluetooth Communications Management

The Bluetooth management software is a module within the Android application that handles all Bluetooth related functions; it is divided into four sub-modules:

- Device Management: Manages the power of the Bluetooth adapter on the Android device.
- Connection Management: Manages connection between the devices.

- Data Transmission: Manages data transfer between the devices.
- Status Monitoring: Monitors and handles Bluetooth events on the Android device.

5.2.1. Device Management

This software module manages the power of the Bluetooth adapter on the Android device. When the application starts, the power status of the Bluetooth adapter will be verified. If Bluetooth is already enabled (powered on), the application will continue to the *main menu*. Otherwise the user will be prompted to give permission to the application to enable Bluetooth. A screenshot of Bluetooth permission request by an application is shown in Figure 7.

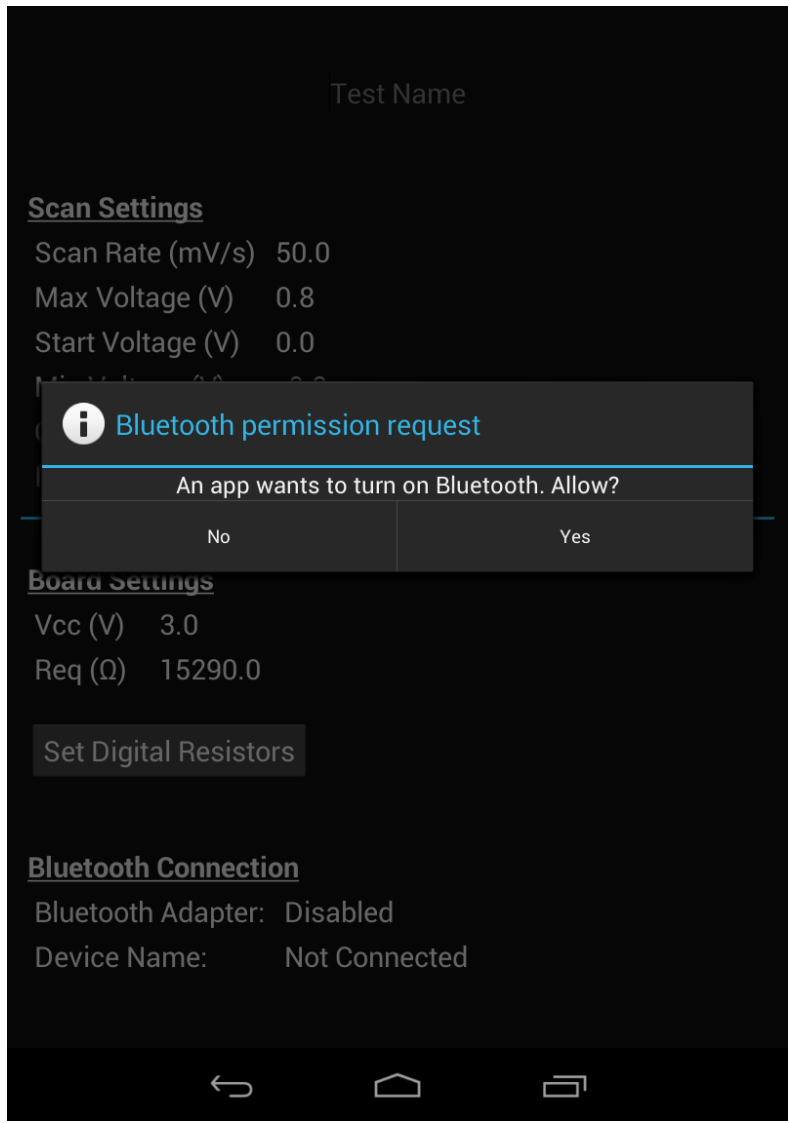

Figure 7. Bluetooth Permission Request.

When permission is granted, Bluetooth will be enabled and the application will continue to the *main menu*. If permission is denied, the application will exit. Figure 8 shows the flow chart describing this process.

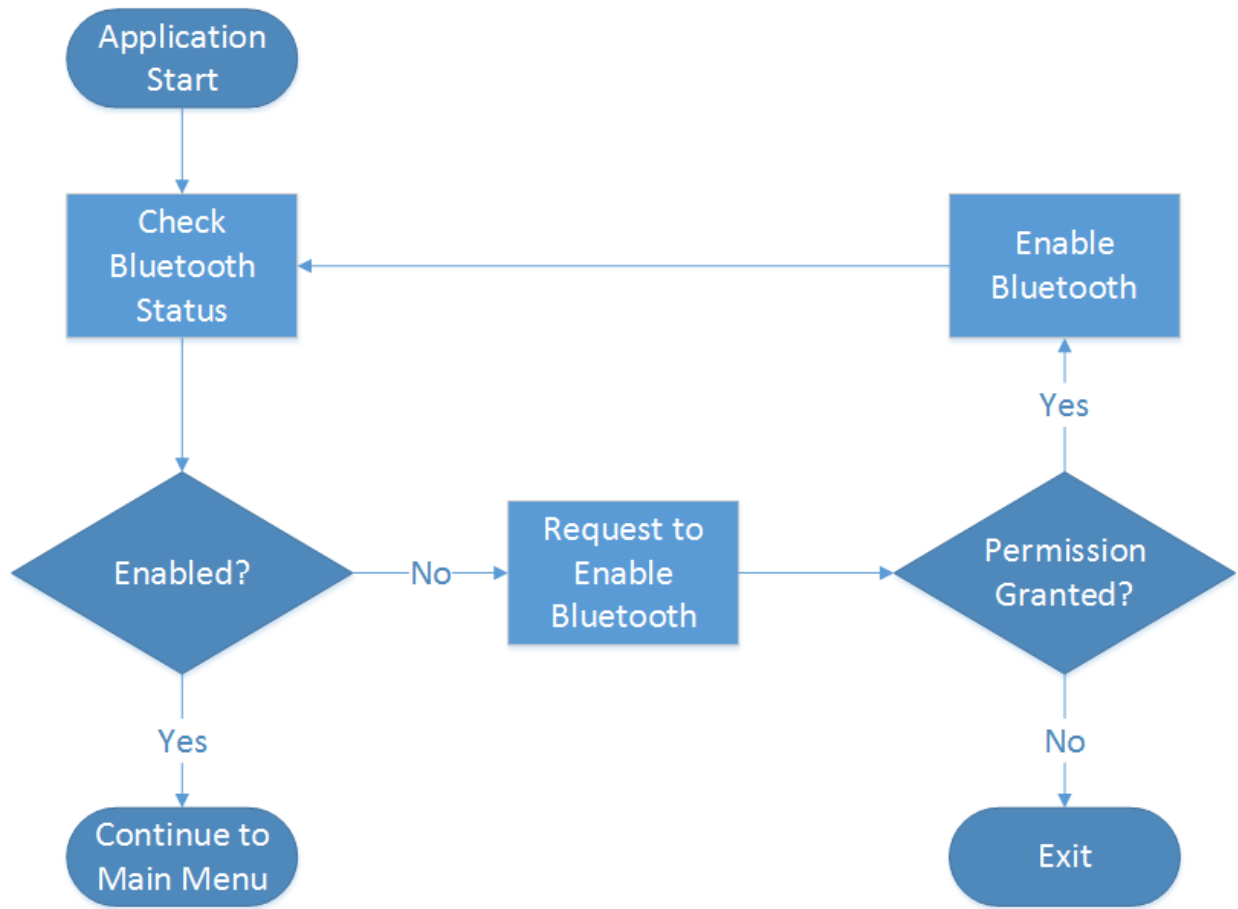

**Figure 8. Bluetooth Enable Check.**

5.2.2. Connection Management

This software module is responsible for establishing and maintaining the Bluetooth connection between the Android and MDI. The Android device and MDI are connected using the standard Android procedure for Bluetooth devices [14]. Figure 9 illustrates a screenshot of a Bluetooth scan result where the Android device has retrieved the names and MAC addresses from the nearby devices. The user chooses the appropriate device to connect with. The prototype shown in Figure 1 has the identifier "ELASH_2996".

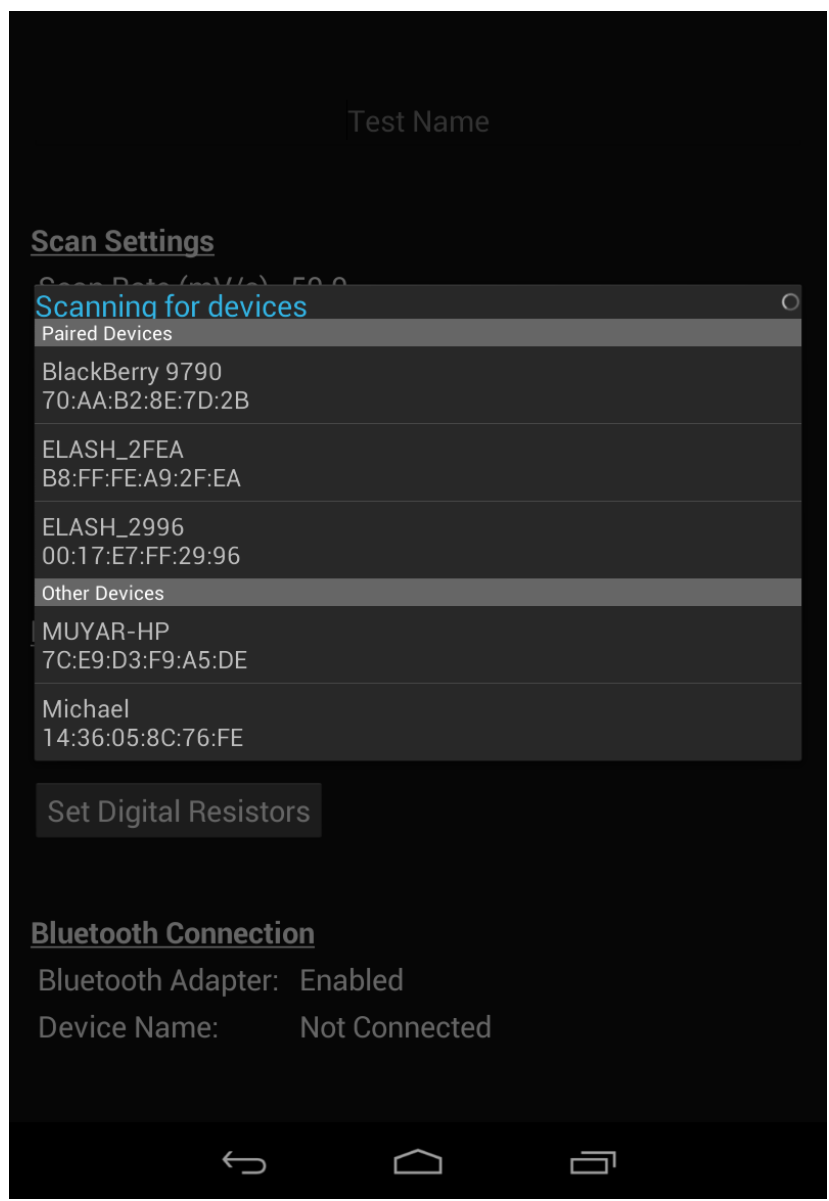

Figure 9. Bluetooth Device Discovery.

When the Android device connects to the MDI for the first time, the user will be prompted to pair the two devices. Pairing allows the Android device to remember the connection [14].

## 6. Functional Verification

The functional performance of the developed prototype was verified by performing a standard electrochemical test using potassium ferricyanide as the sample solution. The results obtained using the prototype were verified by comparing them with the results obtained using the CH

Instruments (CHI 1200B) commercial potentiostat. The electrode unit used for the tests was manufactured by Bioanalytical Systems Inc (BASi). Tests were conducted by immersing the electrodes in the sample solution and measuring the electrochemical response.

*Instruments* - The MDI was used along with several android mobile devices for the electrochemical tests. The tests were carried out on solutions of 1M potassium ferricyanide $K_3[Fe(CN)_6]$. The glassy carbon electrode (GCE) was used as a WE. The Silver/Silver Chloride (Ag/AgCl) with 1M KCl was used as a RE and platinum wire was used as a CE.

*Setup and Procedure* - The WE surface was cleaned by first mechanically polishing it with the BASi polishing kit and cleaning it with acetone and double deionized water. CV was then performed on the WE in 1X Phosphate Buffered Saline (PBS; 8% NaCl, 0.2% KCl, 1.44% $Na_2HPO_4$, 0.24% $KH_2PO_4$) of pH 7.0 between -0.8 to 0.8V with a scan rate of 50mV/s. [7] [15] [16].

*Testing* - For testing and verification, the three electrodes were immersed in the 4ml working volume of the solution and scanned in a potential window of -0.8V to 0.8V at a scan rate of 50mV/s.

*Results and Verification* - The measured data was acquired and plotted. For verification, the same solution was tested using the CHI potentiostat. The result for this test is shown in Figure 10. From the results of the CHI, the peak potentials occur at 0.28V and 0.18V. In the graph obtained using the MDI shown in Figure 6, the peak potentials occur at 0.30 V and 0.13V. It can be observed that the overall trend between the two graphs is the same. There is a discontinuity in the graph because of the limitation of the data acquisition while scanning. This could be improved by fine-tuning the plotting software in future revisions to compensate for the discontinuity. If the results obtained from the commercial potentiostat are used as a standard, it can be observed that the there is a 7.14% and 27.78% error in the peak potential values obtained using the MDI. This error could be attributed to the limitation in the circuitry of the MDI. The accuracy of the MDI could be improved by higher quality circuitry which would help in obtaining better results. These improvements could be implemented into future versions. At this stage, the scope is to successfully demonstrate a proof of concept.

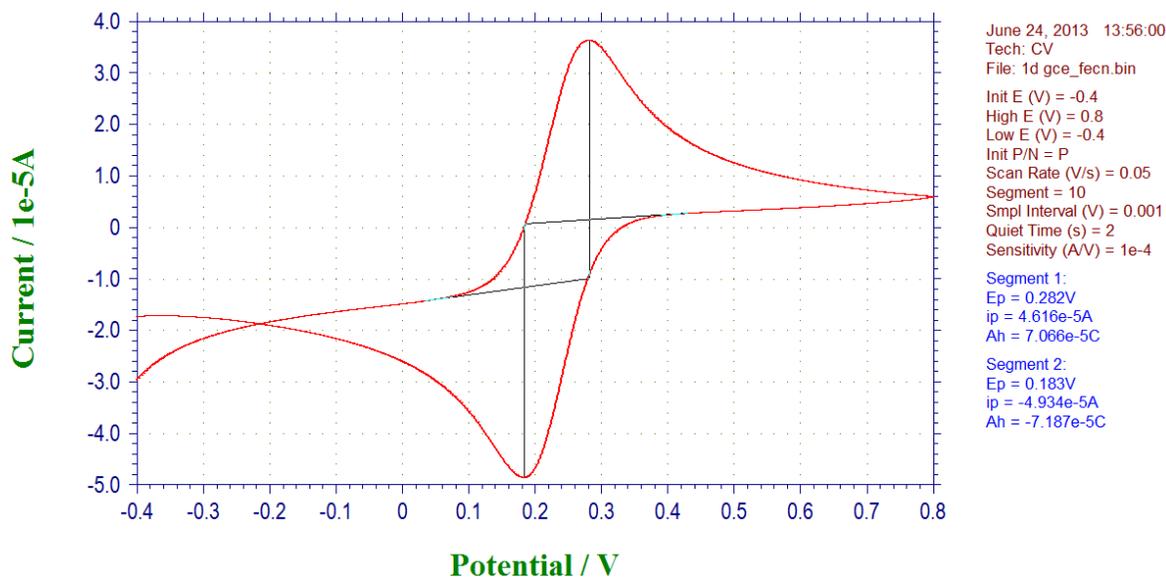

Figure 10. Test result obtained from the CHI potentiostat.

# 7. Conclusion

This work has demonstrated a compact electrochemical system that is suitable for diagnostics. The system consists of a potentiostat interface working along with a mobile device with Bluetooth connectivity. The detected data can be displayed for analysis, and transferred to a centralized healthcare centre for consultation. The portability and low cost of the system allows this technology to be applied for point of care and telemedicine.

# Acknowledgements

This research was supported by the funds from Grand Challenges Canada Program and Natural Sciences, Engineering Research Council, Canada, Shastri Indo-Canadian Institute, Wighton Funds and India Mobility Fund. We gratefully acknowledge the collaboration with the Centre for Biotechnology and Vellore Institute of Technology, India.